\documentclass[prb,twocolumn,showpacs,preprintnumbers,amsmath,amssymb]{revtex4}

\usepackage{graphicx}

\begin{document}


\title{Photoluminescence rings in Corbino disk at quantizing magnetic fields}

\author{V.~Nov\'ak}
\email{vit.novak@fzu.cz}
\author{P.~Svoboda, M.~Cukr,}
\affiliation{%
Institute of Physics AS CR, Cukrovarnick\'a 10,
162~53 Praha 6, Czech Republic}
\author{W. Prettl}
\affiliation{%
Institut f\"ur Experimentelle und Angewandte Physik,
Universit\"at Regensburg, \\93040 Regensburg, Germany}

\date{\today}

\begin{abstract}
Spatially resolved photoluminescence of modulation doped AlGaAs/GaAs 
heterojunction was investigated in a sample of Corbino disk 
geometry subject to strong perpendicular magnetic fields. Significant
spatial modulation of the photoluminescence was observed in
form of one or more concentric rings which travelled across the sample
when the magnetic field strength was varied. A topology of the 
observed structure excludes the possibility of being a trace of
an external current. The effect is attributed to formation of
compressible and incompressible stripes in a 2DEG density 
gradient across the sample.
\end{abstract}


\maketitle

Photoluminescence (PL) of AlGaAs/GaAs het\-ero\-struc\-tures 
contains information on a two-dimensional electron gas (2DEG) 
confined in the system. Various PL features have been identified
typical for 2DEG in strong perpendicular magnetic fields.
\cite{Kuk,Turb,Gold,Couz,Ray,Zin,Shi}
However, little attention has been paid to the {\it spatial}
distribution of the PL. This is a relevant problem especially 
in the case of a 2DEG carrying an unballanced external current
in magnetic fields, when a self-induced spatial inhomogeneity may 
be expected. 
A characteristic current dependent inhomogeneity 
of the PL has indeed been observed in recent experiments.
\cite{Paas,Nov01,Franken}
However, their interpretation is not straightforward: whereas in
Refs.~\onlinecite{Paas,Franken} regions of bright PL along the sample 
edges have been attributed to a locally enhanced hole density, in 
Ref.~\onlinecite{Nov01} a non-radiative decay of excitons due to 
a vertical electric field field has been
made responsible for observed stripes of quenched PL. In neither
of the cases the PL structures could be directly and unquestionably 
associated with a current trace, in contrast e.g.~to PL images of
current filaments in a high-purity bulk GaAs.\cite{Eberle,Hirsch}

In a classical rectangular Hall bar sample it is
difficult to distinguish, whether an observed inhomogeneity
stems primarily from an inhomogeneous longitudinal current, or from
a nonlinear transversal (Hall) potential.
In a sample with rotational symmetry such a distinction should be
possible merely by topological arguments: whereas the potential
distribution preserves the symmetry of the edges, a trace of the 
current flow has to break this symmetry, since it connects the 
contacts of the sample. 
With the above idea in mind we investigated a spatial distribution of
photoluminescence in a 2DEG sample with a Corbino disk geometry.

The investigated sample was cut from a modulation doped single
heterostructure of $\rm Ga_{0.7}Al_{0.3}As/GaAs$ grown by molecular 
beam epitaxy on a semiinsulating substrate. The heterostructure
consisted of $1.5\,\mu$m GaAs buffer layer, 10~nm GaAlAs spacer layer, 
100~nm GaAlAs source layer with nominal doping $10^{18}\,\text{cm}^{-3}$,
and 20~nm GaAs cap layer. On a $5 \times 5$~mm piece of the material
two concentric ring contacts were patterned by annealing of
Au/Ge/Ni. The contacts defined an annulus of 0.85~mm and 1.25~mm
inner and outer radii, respectively. Inside the inner ring a
free circular area has been left floating as a currentless reference.

The sample was mounted
on an insulating carrier, and placed into a superconducting
magnet inside of a liquid helium cryostat. The sample could be 
homogeneously illuminated by a set of red light emitting diodes;
in all of the optical experiments we fed the diodes with a constant
current of 1~mA, corresponding to an estimated power density
of the incident light of 50~$\mu\rm{W/cm^2}$.
The excited photoluminescence was photographed through an optical 
port by a CCD camera, with net resolution of about $8 \mu$m. 
To pick out only the wavelengths of the
near band gap luminescence, an interference filter with 820~nm
central wavelength and $\pm$5~nm bandwidth was put into the optical 
path.

If cooled to 4.2~K while kept in dark, the sample showed pronounced
Shubnikov--de Haas (SdH) oscillations. Using their period together 
with data on the  low-field magnetoresistance we obtained
$3.7\times 10^{15} \text{m}^{-2}$ and $44\;\text{T}^{-1}$ as density
and mobility of the two-dimensional electrons, respectively.
After illumination the density of the 2D electrons rose up 
to $5.7\times 10^{15} \text{m}^{-2}$, accompanied be an increase
in the apparent mobility to $60\;\text{T}^{-1}$. It should be noted,
however, that a strong parallel conductivity evolved at the same
time, accompanied by a deformation of the SdH oscillations and
by a strong reduction of their amplitude.
This parasitic conductivity persisted up to temperature of about 115~K, 
indicating activated DX centers in the ternary layer.\cite{Moon}

\begin{figure}[ht]
\includegraphics[width=1.0\columnwidth]{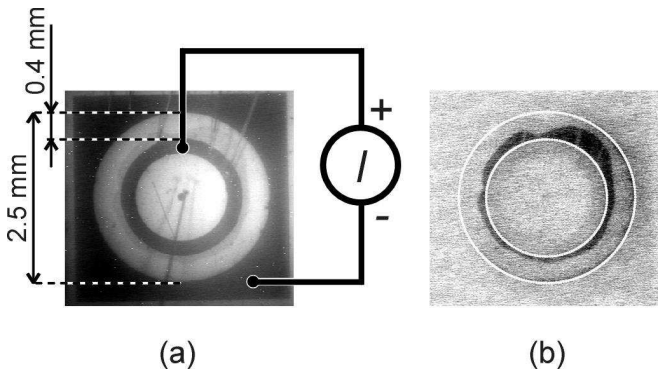}
\caption{
{\bf(a)} PL image of sample with current $I=+200\;\mu$A at $B=3$~T.
PL is seen as light gray, black areas are the metallic contacts.
Central circular area is currentless.
{\bf(b)} Difference of (a) and the PL image of a currentless sample.
Contact edges of the Corbino sample under current (otherwise 
invisible in a difference image) are marked with dotted white lines.
}
\label{1circle}
\end{figure}

Figure \ref{1circle} shows the basic result: when an external
current is applied, a weak spatially inhomogeneous reduction of the PL 
intensity between the contacts occurs (Fig.~\ref{1circle}a). 
The image of the quenched
area can be highlighted by subtracting the constant PL background
of the currentless sample at the same magnetic field. The 
difference image reveals a well defined dark structure in
form of an irregular {\it closed loop}, which encircles the 
inner contact. When reversing the current the dark structure
appears adjacent to the outer contact.
As it will be discussed later, the particular shape,
size and contrast of the quenched structure depend on
the applied current and magnetic field; however, at all
practical conditions the PL structure {\it never} 
appears to connect the two current contacts. Thus, 
a possible filamentation of the imposed current flow may
immediately be excluded as a primary origin of the observed 
PL inhomogeneity. 

The PL pattern is independent of the direction of
the perpendicular magnetic field.
On the other hand, it exhibits an oscillatory
behavior when its strength $B$ is monotonously swept. 
At first, a dark ring 
structure arises adjacent to the positive contact;
on increasing $B$ the dark structure departs from the contact
and travels radially across the sample; with further increase
in $B$ the structure arrives at the negative contact and
disappears; a new dark ring arises at the positive
contact when $B$ is further increased, and the scenario
repeats for each period of SdH oscillations measured on the
sample voltage.

\begin{figure}[h]
\includegraphics[width=1.0\columnwidth]{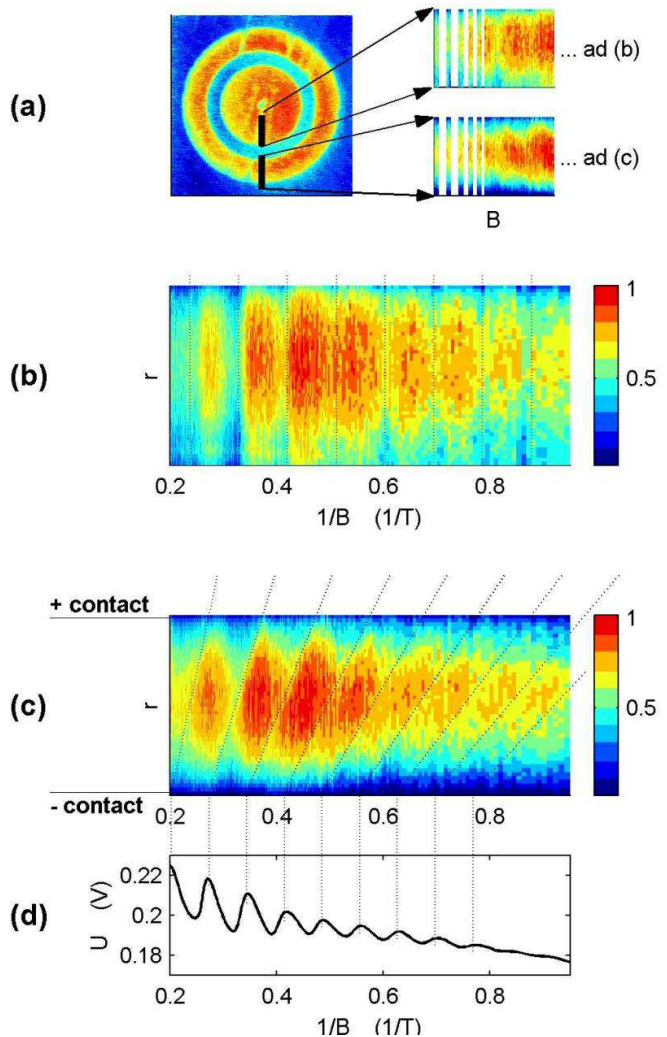}
\caption{
{\bf(a)} Construction of streak images from series of sample PL images
at varying $B$. Black vertical balks indicate sections
in the currentless (upper balk) and in the current biased (bottom balk)
parts of the sample. Vertical dimension of both sections is $400 \mu$m.
{\bf(b)} Streak image of the sample section without current.
Relative PL intensity is color coded according to palette on the right.
{\bf(c)} Streak image of the sample section with current bias
$I=100 \mu$A. The fan of dotted eye-guides indicates the dark structures
(rings) travelling from the positive to the negative contact in increasing
magnetic field, in contrast to standing oscillations in Fig.(b).
{\bf(d)} Electrically measured oscillations of the sample voltage.
Coincidence of the voltage maxima and the dark structure extinctions
at the negative contact is indicated by the vertical dotted lines.
}
\label{streaks}
\end{figure}

The PL rings emerge from a homogeneous PL background approximately 
simultaneously
with the onset of electrically detectable SdH oscillations, similarly 
to the longitudinal quenched stripes in a Hall-bar geometry.\cite{Nov01}
Bad signal-to-noise ratio, especially at low magnetic fields, 
and a weak contrast of the dark structures make it difficult
to determine exactly their contours. 
However, a convincing impression of dark rings repeatedly 
travelling from one contact to another is obtained, if 
a series of static PL images taken at a slowly increasing magnetic 
field is joint together and looked at as a quickly running video 
sequence. To demonstrate this behavior in a single static figure
we extracted a thin radial line
out of each "frame" of such a
sequence, and stacked these line-profiles one after another
in a streak image,
with $1/B$ as abscissa, Fig.~\ref{streaks}(a). 

In the sample without current only vertical bursts of the PL
intensity are seen in the streak image, Fig.~\ref{streaks}(b),
separated by darker grooves of reduced photoluminescence.
The bursts occur simultaneously (though not uniformly) along
the whole radial section, clearly demonstrating {\it standing} 
oscillations of the PL, which correspond to a periodic change 
in the occupation of the highest Landau level. 
If an external voltage is applied,
the dark grooves in the streak image get tilted, Fig.~\ref{streaks}(c), 
reflecting a {\it motion} of the dark rings in varying $B$. 
When reversing the current bias, also the travelling direction 
(i.e.~the tilting angle of the grooves in the streak image) reverses, 
pointing always from the positive to the negative contact in 
increasing $B$.

Locations of the dark grooves in $1/B$ can well be approximated by
straight lines, as indicated by a set of eye-guides in
Fig.~\ref{streaks}(b,c). This is in a clear contrast with
a steep dependency of edge channel positions, theoretically
treated by Ref.~\onlinecite{Chklo} and confirmed experimentally
for low currents e.g.~in Ref.~\onlinecite{AhlsCorb}.
The travelling velocity of the PL rings (i.e.~the tilting angle
of the dark grooves in Fig.~\ref{streaks}(c)) obviously changes 
from one SdH period to another, increasing with a decreasing 
filling factor. Accordingly, two or more rings can be observed 
simultaneously at low magnetic fields, whereas 
only one ring at a time exists at high $B$. 

\begin{figure}[h]
\includegraphics[width=1.0\columnwidth]{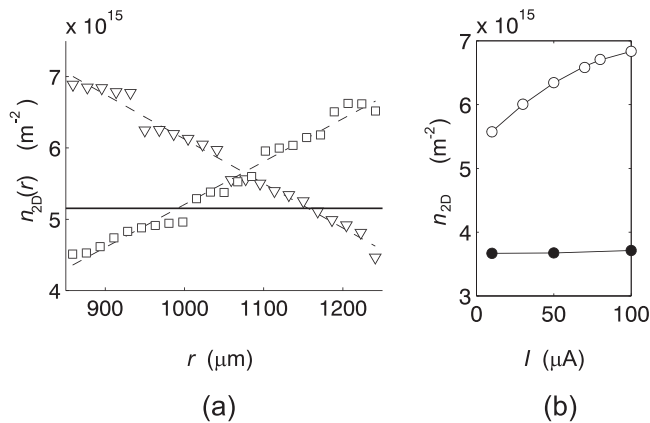}
\caption{
{\bf(a)} 2DEG density profiles evaluated from the $1/B$ periodicity
of the streak image in Fig.~\protect\ref{streaks}c. Triangles and squares correspond
to $I=+100\;\mu$A and $I=-100\;\mu$A, respectively, with higher $n_{2\rm{D}}$
always at the negative electrode. Solid line indicates the zero-current 2DEG
density acquired from Fig.~\protect\ref{streaks}b.
{\bf(b)} 2DEG density evaluated from oscillations of 
sample voltage. Full and open circles correspond to sample
before and after illumination, respectively. 
}
\label{grad}
\end{figure}

The $1/B$-period of the PL ring occurrence (i.e.~the dark grooves
spacing) changes with the radius $r$, as also seen from
Fig.~\ref{streaks}(c). 
On the other hand, at each radial coordinate $r$ 
the PL image period in $1/B$ is fairly constant in the
whole range of $B$. Thus, using this period dependence, 
a two-dimensional electron density $n_{2\rm{D}}(r)$ can be extracted 
from the streak image Fig.~\ref{streaks}(c) as a function
of the radial coordinate.
The result shown in Fig.~\ref{grad}(a) suggests that 
a significant gradient of the electron density exists in 
a sample with an externally imposed current. 
A slope of the gradient depends on the current magnitude,
maximum $n_{2\rm{D}}$ is always at the negative contact.

Under a constant current bias the sample voltage exhibits well 
pronounced oscillations in varying $B$. Their periodicity coincides 
with that of the dark PL ring recurrence at the negative
contact, as seen in Figs.~\ref{streaks}(c,d).
Accordingly, also the voltage oscillation period varies with 
the current bias, reflecting a local filling factor near
the negative contact. A current dependency of the electron 
density at the negative contact (i.e.~the maximum $n_{2\rm{D}}$),
acquired from the voltage period, is plotted in Fig.~\ref{grad}b. 
It should be stressed that no such dependence can
be found in a sample prior to the first illumination, where
the period of the voltage oscillations remains constant even 
for currents exceeding $100 \mu$A.

\begin{figure}[h]
\includegraphics[width=1.0\columnwidth]{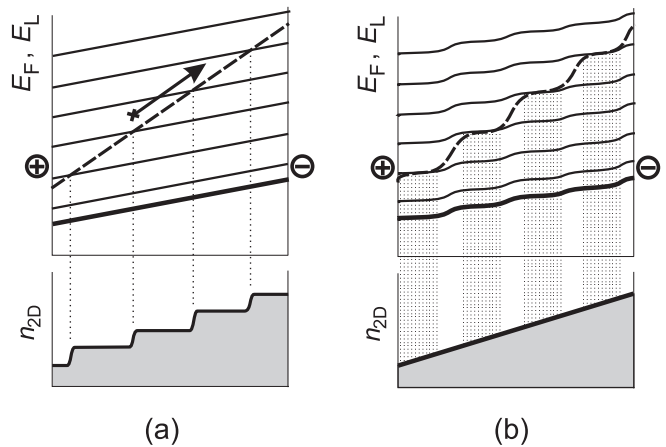}
\caption{
(a) Schematic sketch of Fermi (dashed line) and Landau levels 
(solid lines) in one electron approximation. Arrows indicate 
direction of crossing point motion in increasing $B$.
(b) Sketch of Fermi and Landau levels with electron screening.
Compressible stripes are indicated by dotted balks.
}
\label{levels2}
\end{figure}

A qualitative explanation of the observed PL structures
can be constructed starting from a one electron picture.
Generally, in an electrically biased sample the slopes of both 
the Landau levels and the Fermi level have the same signs,
given by the bias polarity. Furthermore, in a sample with
a finite concentration gradient pointing from the positive
to the negative contact  
the Fermi level crosses the Landau level system as
schematically shown in Fig.~\ref{levels2}(a). Upon
increasing $B$, the spacing of the Landau levels
grows and the crossing points of the Landau levels and
the Fermi level shift toward the negative contact.
Let us note that this direction agrees with the sense
of the observed PL ring motion and that this coincidence
would not occur in case of an opposite concentration
gradient.

In a more realistic approximation, the effect of Fermi
level pinning has to be taken into account:\cite{Guven}
the electrons close to the Fermi surface tend to screen 
an external electric field and form compressible stripes 
of finite widths.\cite{Chklo} 
Accordingly, instead of a set of {\it lines}, a set of 
compressible {\it stripes} of partially
occupied Landau levels arises, separated by
narrow incompressible regions of fully occupied or totally 
empty states, Fig.~\ref{levels2}(b).
The sense of the stripe motion in varying $B$ remains
the same as in the one electron picture, i.e.~the 
experimentally observed one.

A photoluminescence involving the 2D electrons depends
on the magnetic field both with its wavelength 
and its intensity. However, the observed spatial structures
in the PL images are to be attributed solely to changes in 
the intensity: 
except for a contrast, the same spatial structures have been
obtained using interference filters of 820 and 830~nm 
central wavelengths. A plausible mechanism of the
PL intensity variation can again be found within the 
qualitative picture according to Fig.~\ref{levels2}(b):
the PL intensity exhibits a maximum for electrons
from the Fermi surface (Fermi edge singularity),
i.e.~from the compressible stripes in our case, whereas 
the PL is reduced in the incompressible ones.

A question remains on the origin of the gradient of
the 2D electron concentration. In fact, for a Corbino
sample geometry under external current an inhomogeneous 
2DEG density has been 
predicted theoretically.\cite{Dyak,Shik1,Shik2}
There are, however, reasons for which we
believe that those theories are not fully pertinent to 
our experiment. In particular, we have observed differences
in the behavior of the sample prior to and after
the illumination. For the most significant one we hold 
the above mentioned change in the electrically
measurable SdH oscillations in the sample voltage:
whereas there was no detectable current induced change 
in their period before the first illumination, 
a clear dependence has evolved between the period
and the sample current after the illumination;
let us recall that this dependence obviously correlates 
with the periodicity of the observed PL structures.
Furthermore, a transition from a perfectly linear to
a more complex and $B$-dependent form of sample's
I-V curve has been found during the illumination.

As a more feasible explanation we regard a model which assumes
a presence of some conducting (either contacted or floating)
layer parallel to the 2DEG. The existence of such a parallel
layer has been suggested by several authors and indirectly 
supported by various experimental 
results.\cite{Coldien,Knot,Jian,Zheng,Ebe,Nov01}
Possible origins include photogenerated holes accumulated
at the buffer-substrate interface, DX-center induced
conductivity in the ternary layer, and others.
By solving an electrostatic problem corresponding to
such a double-layer system, an 
electron density gradient can be shown to arise 
if a non-zero potential drop exists between the
banks of the 2DEG layer.\cite{Coldien, Shask1}
A self-consistent solution of the electrostatics,
Fermi statistics of 2D-electrons, and a realistic
longitudinal conductivity function $\sigma_{xx}$
would then yield a quantitative correction to 
Fig.~\ref{levels2}(b).

{\it Conclusions.} The PL pattern observed in a current 
biased sample at 
quantizing magnetic field has the form of one or more
rings concentric with the contacts of the Corbino disk.
This topology excludes the possibility of the PL
pattern being a trace of a channel carrying an
unballanced external current. 

The PL rings have been found to travel periodically across 
the sample when sweeping the magnetic field. 
The rings travel the whole path from the positive to the negative
contact in increasing magnetic field, with a velocity differing 
qualitatively from the one expected for the edge channels in a 
confining potential in a homogeneous sample.

Indications have been found that the sample behavior 
significantly differs before and after the first illumination. 
We suppose that the illumination---besides building up a 
parasitic conductivity---leads to a formation of a persistent 
equipotential layer parallel to the 2DEG. The existence of such
a layer in a sample under current gives rise to an electron 
density gradient, confirmed experimentally by the varying
$1/B$ periodicity of the observed photoluminescence pattern.
Coming out from the non-zero density gradient a qualitative
model has been formulated, which explains the observed
PL structures in terms of compressible and incompressible
stripes.

\begin{acknowledgments}
The authors thank K.~V\'yborn{\'y} for valuable discussions, 
and Z.~V\'yborn\'y and V.~Jurka for technical assistance. 
Support by the Humboldt foundation and the Deutsche 
Forschungsgemeinschaft is gratefully acknowledged 
by V.N. and W.P. The work was done in the framework of 
AV0Z1-010-914 program.
\end{acknowledgments}

\end{document}